\documentclass{moriond}

\def\be{\begin{equation}}
\def\ee{\end{equation}}
\def\bea{\begin{eqnarray}}
\def\eea{\end{eqnarray}}

\renewcommand{\vec}[1]{\mathbf{#1}}

\def\chieff{\chi_\mathrm{eff}}
\def\chieffpm{\chi_{\mathrm{eff}\pm}}
\def\chieffp{\chi_{\mathrm{eff}+}}
\def\chieffm{\chi_{\mathrm{eff}-}}
\def\rudpm{r_{\mathrm{ud}\pm}}
\def\rudp{r_{\mathrm{ud}+}}
\def\rudm{r_{\mathrm{ud}-}}
\def\thetapert{\theta_\mathrm{pert}}
\def\risco{r_\mathrm{ISCO}}
\def\rligo{r_\mathrm{LIGO}}

\newcommand{\Photo}{}

\begin{document}

\vspace*{4cm}

\title{Unstable binary black-hole spins: post-Newtonian theory and numerical relativity}

\author{Matthew Mould}

\address{School of Physics and Astronomy \& Institute for Gravitational Wave Astronomy,\\University of Birmingham, Birmingham, B15 2TT, United Kingdom}

\maketitle\abstracts{
Spin precession occurs in binary black holes whose spins are misaligned with the orbital angular momentum. Otherwise, the spin configuration is constant and the subsequent binary dynamics and gravitational-wave emission are much simpler. We summarize a series of works which has shown that, while three of the aligned configurations are stable equilibria, the `up-down' configuration, in which the heavier (lighter) black hole is (anti) aligned with the orbital angular momentum, is unstable when perturbed; at a critical point in the inspiral the black hole spins begin to tilt wildly as precession takes over. We present two equivalent approaches to derive the instability onset based on multitimescale post-Newtonian techniques, and point out that the instability has a predictable endpoint. Finally, we demonstrate the presence of this precessional instability in the strong-field regime of numerical relativity with simulations of aligned-spin binaries lasting $\sim100$ orbits before merger. The spins of up-down systems can tilt by $\sim90^\circ$, leaving a notable imprint in the emitted gravitational-wave signals and providing a possible mechanism to form precessing systems in astrophysical environments from which sources are preferentially born with (anti) aligned spins.
}

\section{Introduction}
\label{section:intro}

Binary black-hole (BBH) spins are important astrophysical observables in gravitational-wave (GW) astronomy. Their dynamics require detailed modeling in order to produce accurate waveforms\cite{Apostolatos:1994mx} and are crucial for determining the formation channels of GW events.\cite{Mandel:2018hfr} Typically, the directions of the orbital plane and BH spins change over an inspiral due to general-relativistic spin precession,\cite{Apostolatos:1994mx,Kidder:1995zr} but not when the spins are aligned with the orbital angular momentum. For unequal-mass binaries, there are four such alignments. We label a BH with aligned (antialigned) spin as `up' (`down'), and thus a binary is labeled as, e.g., `up-down', where the direction before (after) the hyphen refers to the primary (secondary) BH. We demonstrate the existence of a precessional instability in up-down BBHs\cite{Gerosa:2015hba,Lousto:2016nlp} and find its asymptotic endpoint\cite{Mould:2020cgc} with post-Newtonian (PN) techniques, before also demonstrating its effect on the spin dynamics and GW emission of inspiraling BBHs in numerical relativity (NR) simulations.\cite{Varma:2020bon}

\section{Unstable spin precession}
\label{section:pn}

We use geometric units $G=c=1$. A binary of spinning BHs is described by component masses $m_1>m_2$ (or the mass ratio $q=m_2/m_1$ and total mass $M=m_1+m_2$) and its angular momenta --that of the orbit, $\vec{L}$, and of the BH spins, $\vec{S}_1$ and $\vec{S}_2$. They can be described by six parameters: their magnitudes $L$, $S_1$ and $S_2$, the spin tilt angles $\theta_i = \arccos[\vec{S}_i\cdot\vec{L}/(S_iL)]$ ($i=1,2$), and the azimuthal angle $\Delta\Phi$ between the in-plane spin components. Since the spin magnitudes (or the dimensionless spins $\chi_i=S_i/m_i^2$) are conserved, just $(L, \theta_1, \theta_2, \Delta\Phi)$ are free parameters. Reparameterizing by $(L,\theta_1,\theta_2,\Delta\Phi) \mapsto (L,J,\chieff,S)$, $L=m_1m_2\sqrt{r/M}$ and the magnitude of the total angular momentum $J=|\vec{L}+\vec{S}_1+\vec{S}_2|$ vary only on the longer radiation-reaction timescale, on which the orbital separation $r$ shrinks, while the effective aligned spin, $\chieff=(\chi_1\cos\theta_1+q\chi_2\cos\theta_2)/(1+q)$,
is conserved.\cite{Racine:2008qv} Therefore, on the shorter precession timescale, the BBH spins are described by a single time-varying parameter: the total spin, $S=|\vec{S}_1+\vec{S}_2|$.\cite{Kesden:2014sla,Gerosa:2015tea}

\subsection{Effective potential approach}
\label{section:potential}

Since $\chieff$ is conserved, on the precession timescale $S$ is constrained to move between extrema $S_-\leq S\leq S_+$ in the $S$--$\chieff$ plane. Here, $S_\pm$ are the solutions of $\chieff=\chieffpm(S)$, where the curves $\chieffpm(S)$ act as effective potentials for spin precession and form a closed loop.\cite{Kesden:2014sla,Gerosa:2015tea}

In the mutually aligned up-up and down-down configurations, this loop reduces to a single point and no spin precession can occur. In the antialigned configurations, oscillations of $S$ are possible except at the top and bottom of the loop, where $\chieff$ is extremized. Antialignment implies $S=|S_1-S_2|$ (the leftmost point on the loop), where $\chieffp(S)=\chieffm(S)$. The down-up configuration sits at the maximum of the loop where no spin precession can occur, since $d\chieffpm/dS<0$. For the up-down configuration, $d\chieffpm/dS>0$ at large orbital separations, so that it is located at the minimum. However, for orbital separations $\rudp>r>\rudm$, where
\be
\rudpm = \frac{(\sqrt{\chi_1} \pm \sqrt{q\chi_2})^4}{(1-q)^2}
\, ,
\label{eq:rudpm}
\ee
$d\chieffm/dS<0$ and $S$ can oscillate. Below $\rudm$, $\chieffp(S)$ also changes sign so that the up-down configuration is now located at the maximum of the loop. In the intermediary range $\rudp>r>\rudm$, the up-down configuration is an unstable fixed point.\cite{Gerosa:2015hba}

\subsection{Harmonic oscillator approach}
\label{section:harmonic}

An equivalent approach is to consider the time-dependence of $S$ which, using the 2PN orbit-averaged spin precession equations,\cite{Racine:2008qv} is determined on the precession timescale by $(dS^2/dt)^2 \propto (S_+^2-S^2) (S^2-S_-^2) (S^2 - S_3^2)$, where $S_3$ is a third nonphysical root.\cite{Chatziioannou:2017tdw} For an aligned spin configuration denoted by $S_*$, the total spin is a constant, $S_*^2=S_-^2=S_+^2$. A perturbation $S^2-S_*^2$ to the aligned spin configuration gives the leading order second time derivative
\be
\frac{d^2}{dt^2}(S^2-S_*^2) + \omega^2(S^2-S_*^2) \approx 0
\, , \quad
\omega^2 \propto 3S_*^2 - S_+^2 - S_-^2 - S_3^2
\, .
\label{eq:omega}
\ee
Equation~\ref{eq:omega} is that of a harmonic oscillator: when $\omega^2>0$, the perturbations cause small amplitude oscillations of the BH spins around the exact alignment; when $\omega^2<0$, the perturbed system is unstable and precession leads to large spin tilts. Stability transitions, when $\omega=0$, occur only for up-down binaries at $r=\rudpm$, in precise agreement with the previous approach.\cite{Mould:2020cgc}

Numerical exploration of inspiraling BBHs\cite{Gerosa:2016sys} revealed that the precessional instability develops over a typical decrease $\delta r\approx25M$ in the orbital separation and that, at small separations, the binaries cluster at well-defined endpoints. At large separations, up-down binaries correspond to the $\Delta\Phi=0$ category of resonant configurations\cite{Schnittman:2004vq,Gerosa:2015hba} and, during the post-instability inspiral, tend to be recaptured into this resonance. By locating analytically the location of this resonance in the small separation limit, we find that, for the unstable up-down BBHs,
\be
\cos\theta_1 \, , \, \cos\theta_2 \to \frac{\chi_1-q\chi_2}{\chi_1+q\chi_2}
\, , \quad
\Delta\Phi \to 0
\quad
\mathrm{as} \quad r \to 0
\, ,
\label{eq:limit}
\ee
i.e., the spins coalign but become equally misaligned with the orbital angular momentum.\cite{Mould:2020cgc}

\begin{figure}
\centering
\includegraphics[width=1.0\linewidth]{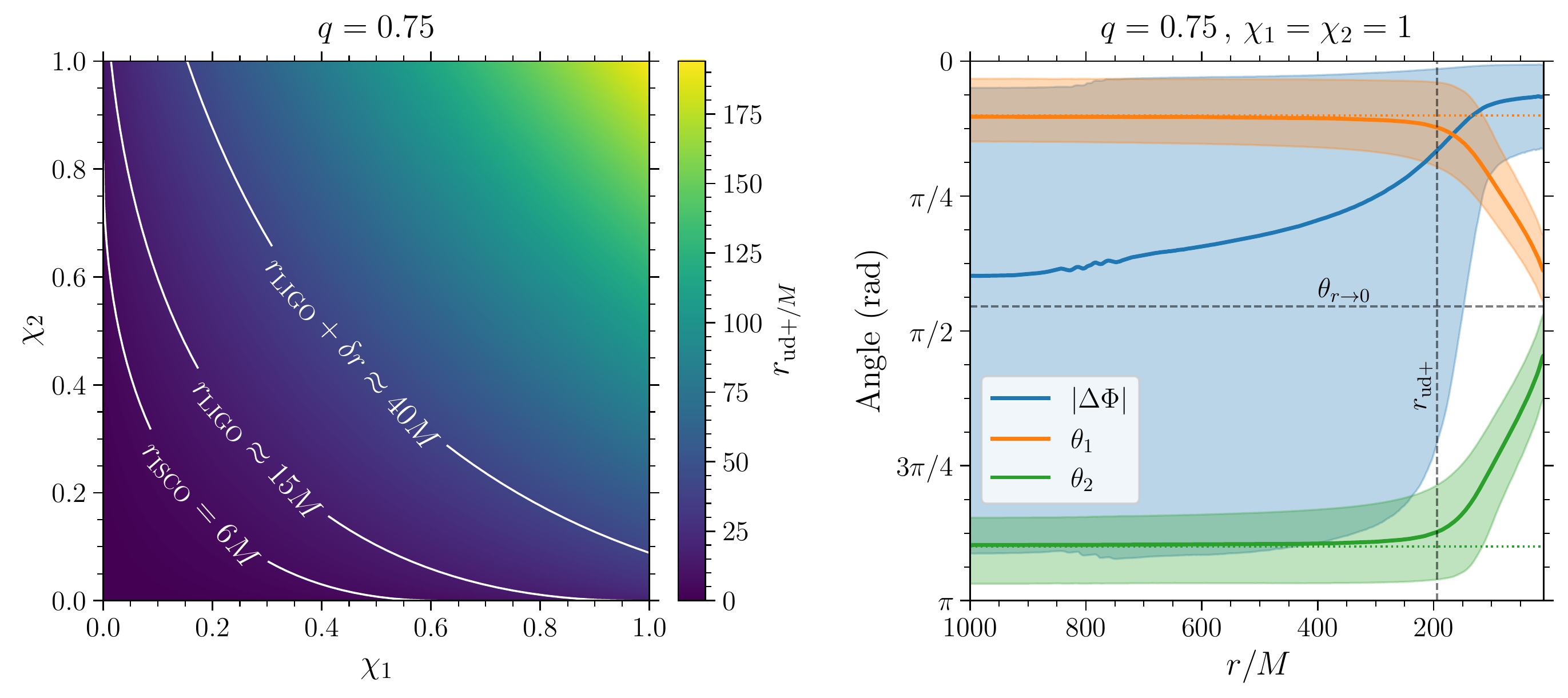}
\caption{Left: Values of the instability onset $\rudp$ for up-down BBHs with mass ratio $q=0.75$ as a function of the dimensionless spins, $\chi_1$ and $\chi_2$. Highlighted for reference (white curves) are the locations of the Schwarzchild innermost stable circular orbit (ISCO), $\risco$, the orbital separation corresponding to a GW frequency of $20~\mathrm{Hz}$ (for total mass $M=50M_\odot$), $\rligo$, and the value of $\rudp$ required for the precessional instability to develop appreciably before a BBH enters the LIGO band, $\rligo+\delta r$. Right: Medians (solid lines) and $90\%$ confidence regions (shaded areas) for evolutions of the spin angles $\theta_1$ (orange), $\theta_2$ (green) and $\Delta\Phi$ (blue) for a population of up-down binaries with $q=0.75$, $\chi_1=\chi_2=1$ and initial spins distributed isotropically within $\approx20^\circ$ of the exact alignment. The vertical dashed grey line shows the location of the instability onset, $r=\rudp$, and the horizontal dashed grey line shows the analytic limit of the tilt angles.}
\label{fig:pn}
\end{figure}

The up-down instability occurs earlier in the inspiral for BBHs with higher spins, as demonstrated in the left panel of Fig.~\ref{fig:pn}. On the other hand, in the limits of equal or very unequal masses the up-down configuration becomes stable. The right panel of Fig.~\ref{fig:pn} displays the evolution of the spin angles for a population of binaries initially close to up-down alignment. The spin tilts remain constant until encountering the instability at $r=\rudp$, from whence they tend to the asymptotic value given by Eq.~\ref{eq:limit} while converging to the $\Delta\Phi=0$ resonance.

\section{Up-down binaries in numerical relativity}
\label{section:nr}

The predictions of Section~\ref{section:pn} are made with approximate PN techniques. To study BBHs in full General Relativity (GR), one must rely on NR simulations. Using the Spectral Einstein Code (SpEC)\cite{Boyle:2019kee} developed by the Simulating eXtreme Spacetimes (SXS) Collaboration, we simulate the late inspirals and mergers of up-down binaries with mass ratio $q=0.9$ and dimensionless spins $\chi_1=\chi_2=0.8$. The binaries are evolved from an initial separation $r=30M$, which is well within the instability regime ($\rudp\approx900M$ and $\rudm\sim10^{-4}$). As controls, we also simulate the stable configurations. In all cases, we perturb the exact alignments by $\thetapert=1^\circ,5^\circ,10^\circ$. The 12 new simulations\cite{Varma:2020bon} (which have been assigned identifiers SXS:BBH:2313--2324) are among the longest in the SXS catalog,\cite{Boyle:2019kee} with each lasting $\approx100$ orbits before merger.

\begin{figure}
\centering
\includegraphics[width=1.0\linewidth]{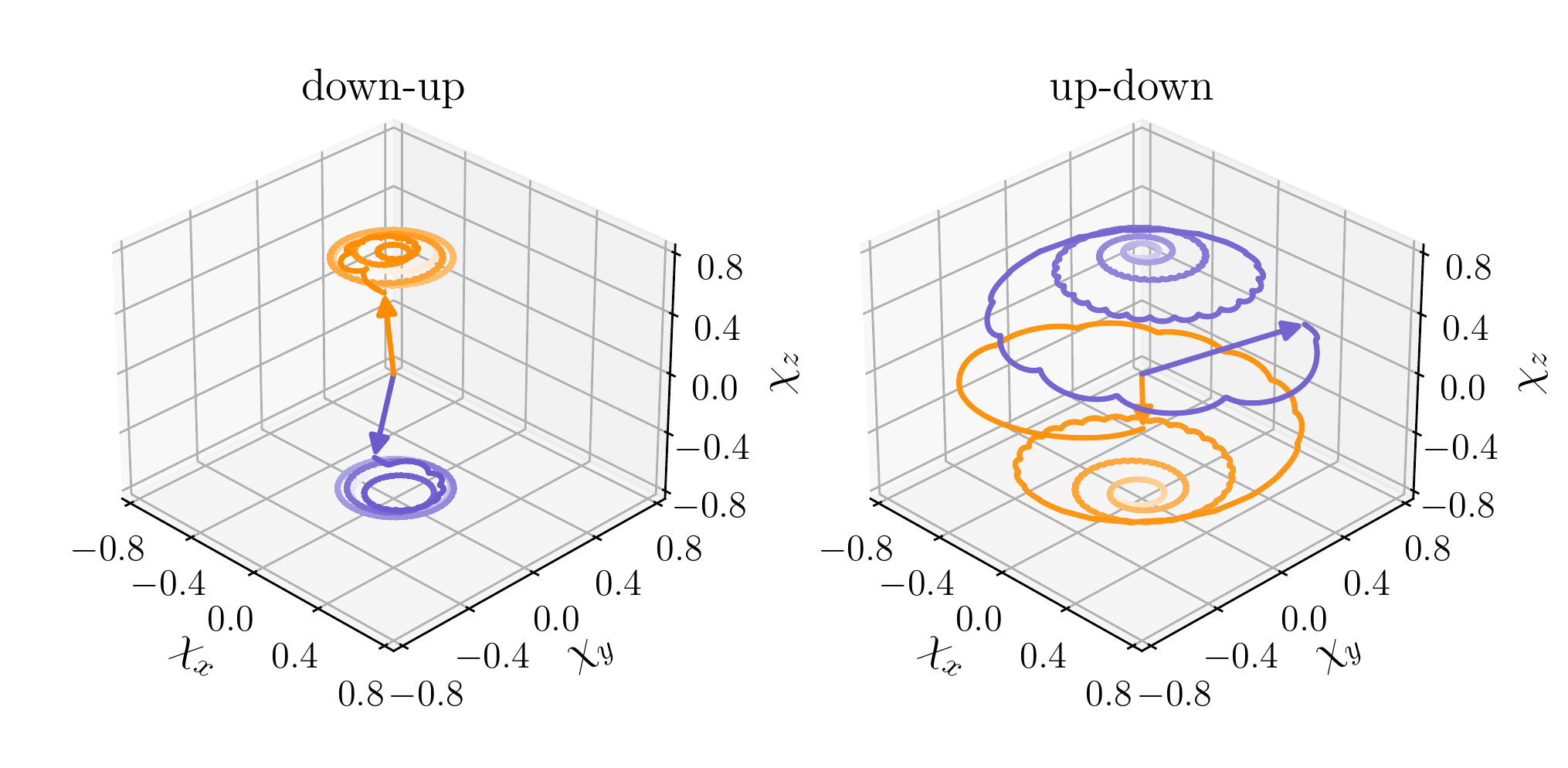}
\caption{Evolutions of the primary (purple) and secondary (orange) BH spins from a NR simulation of an up-down binary (right; SXS:BBH:2324), and for comparison those of a down-up binary (left; SXS:BBH:2323), each with mass ratio $q=0.9$ and dimensionless spins $\chi_1=\chi_2=0.8$. The binaries are initialized with spins perturbed by $10^\circ$ from the exact alignments. The spin vectors are viewed in a frame whose $z$-axis is aligned with the initial direction of the orbital angular momentum. While the down-up binary always remains close to its initial configuration as the component BHs inspiral, the precessional instability causes spin precession in the up-down case, leading to large spin misalignments at merger.}
\label{fig:nr}
\end{figure}

These simulations confirm the existence of the up-down instability in the strong-field regime of GR. Its effect is visualized in Fig.~\ref{fig:nr} for $\thetapert=10^\circ$. While the BH spins are initially close to the exact alignment, close to merger they have tilted wildly, with misalignments reaching up to $\approx90^\circ$; for the control cases, this unstable precessional motion does not occur. The corresponding orbital plane precession causes a transfer of power in the emitted GW signal from the dominant (2, 2) mode to the subdominant (2, 1) mode, which features a characteristic growth in amplitude.

\section{Conclusions}

As predicted with multi-timescale PN techniques\cite{Gerosa:2015hba,Lousto:2016nlp,Mould:2020cgc} and verified by NR simulations,\cite{Varma:2020bon} the spins of up-down BBHs are unstable. If aligned-spin binaries form astrophysically at large orbital separations, the precessional instability will cause a dearth of detected BBH mergers with up-down spins. Instead, they will be present in the detection bands of ground- and spaced-based of GW interferometers with spin tilts close to those given by Eq.~\ref{eq:limit}. Instability-induced precession of the orbital plane is expected to leave an imprint in the GW emission of these systems. As the measurement accuracy of BBH parameter estimation improves, the detection of a precessing GW signal with spin configuration matching that of the up-down endpoint may allow us to infer the occurrence of the precessional instability.

\section*{Acknowledgments}

M. M. is supported by European Union's H2020 ERC Starting Grant No. 945155--GWmining and Royal Society Grant No. RGS-R2-202004.


\section*{References}

\begin{thebibliography}{99}

\bibitem{Apostolatos:1994mx}
T.~A.~Apostolatos {\em et al.}, PRD \textbf{49}, 6274-6297 (1994).

\bibitem{Mandel:2018hfr}
I.~Mandel and A.~Farmer, arXiv:1806.05820 [astro-ph.HE].

\bibitem{Kidder:1995zr}
L.~E.~Kidder, PRD \textbf{52}, 821-847 (1995).

\bibitem{Gerosa:2015hba}
D.~Gerosa {\em et al.}, PRL. \textbf{115}, 141102 (2015).

\bibitem{Lousto:2016nlp}
C.~O.~Lousto and J.~Healy, PRD \textbf{93}, 124074 (2016).

\bibitem{Mould:2020cgc}
M.~Mould and D.~Gerosa, PRD \textbf{101}, 124037 (2020).

\bibitem{Varma:2020bon}
V.~Varma {\em et al.}, PRD \textbf{103}, 064003 (2021).

\bibitem{Racine:2008qv}
E.~Racine, PRD \textbf{78}, 044021 (2008).

\bibitem{Kesden:2014sla}
M.~Kesden {\em et al.}, PRL. \textbf{114}, 081103 (2015).

\bibitem{Gerosa:2015tea}
D.~Gerosa {\em et al.}, PRD \textbf{92}, 064016 (2015).

\bibitem{Chatziioannou:2017tdw}
K.~Chatziioannou {\em et al.}, PRD \textbf{95}, 104004 (2017).

\bibitem{Gerosa:2016sys}
D.~Gerosa and M.~Kesden, PRD \textbf{93}, 124066 (2016).

\bibitem{Schnittman:2004vq}
J.~D.~Schnittman, PRD \textbf{70}, 124020 (2004).

\bibitem{Boyle:2019kee}
M.~Boyle {\em et al.}, CQG \textbf{36}, 195006 (2019).

\end{thebibliography}
\end{document}